\documentclass[
 reprint,
 amsmath,amssymb,
 aps,
]{revtex4}
\usepackage[section]{placeins}
\let\Oldsection\section
\renewcommand{\section}{\FloatBarrier\Oldsection}

\let\Oldsubsection\subsection
\renewcommand{\subsection}{\FloatBarrier\Oldsubsection}

\let\Oldsubsubsection\subsubsection
\renewcommand{\subsubsection}{\FloatBarrier\Oldsubsubsection}
\usepackage{graphicx}
\usepackage{dcolumn}
\usepackage{bm}
\makeatletter
\AtBeginDocument{%
  \expandafter\renewcommand\expandafter\subsection\expandafter{%
    \expandafter\@fb@secFB\subsection
  }%
}
\makeatother

\begin{document}

\title{Link between Alhassid-Levine  and Hioe-Eberly formalisms of $SU(N)$ equation of motion}

\author{Dawit Hiluf}
\email{dawit@post.bgu.ac.il}
\affiliation{Ben-Gurion University of Negev department of Physical Chemistry, Be'er-Sheva 84105, Israel}
\affiliation{Physics Department, Mekelle University, P.O.Box 231, Mekelle, Ethiopia.}
\date{\today}
\begin{abstract}
Geometric representations of solutions provides intuitive physical insights. To which end studying dynamics of Quantum systems via $su (n)$ Lie algebra proves to be convenient way of obtaining geometric solution. In this paper link is established between two formalisms that made use of Lie algebra to describe equation of motion for quantum system. In both approaches  the Hamiltonian and the density matrix are  expressed as a linear combination of the Lie group. To exemplify the approach we consider a very well studied two level system coupled by a laser pulse. Beyond establishing link between these two formalism we obtained two constants of motion by assuming time dependent detuning whose time profile is assumed to be same as the laser pulse. Consequently we have shown how one can have two disjoint subspaces whose evolution vector is independent of each other. 

\end{abstract}

\maketitle



\section{Introduction}
Given any physical system, one can perform certain "operations" or "transformations" with it, example rotations, translations, scale transformations, conformal transformations, Lorentz transformations. Physical transformations form a group from the mathematical viewpoint. In this paper, we make use of the Lie groups $U(n)$ and $SU(n)$, and their respective Lie algebras, generally denoted by $u(n)$ and $su(n)$. The unitary group is defined by: $U(n)\equiv\left\{ U\in M_{n\times n}(\mathbb{C})/UU^+=U^+U=I\right\}$. The special unitary group is defined by: $SU(n)\equiv\left\{ U\in M_{n\times n}(\mathbb{C})/UU^+=U^+U=I,\det (U)=1\right\}$. The group operation is the usual matrix multiplication. We note also that $U(n)$ and $SU(n)$ are compact Lie groups. As a consequence, they have unitary, finite dimensional and irreducible representations. Moreover $U(n)$ and $SU(n)$ are subgroups of $U(m)$ if $m\geq n$. The unitary group has $n^2$ parameters (its "dimension") whereas the special unitary group has $n^2-1$ free parameters (its "dimension"). This means that the unitary group has a Lie algebra generated by the space of $n^2$ dimensional complex matrices while the special unitary group has a Lie algebra generated by the $n^2-1$ dimensional space of hermitian $n\times n$ traceless matrices. 

A very familiar group in quantum mechanics community is the case $n=2$, this is an important group in physics. It appears in many contexts: angular momentum (both classical and quantum), the rotation group, spinors, quantum information theory, spin networks and black holes, the Standard Model, and many other places. The number of parameters of $SU(2)$ is equal to 3. As generators of the Lie algebra associated to this Lie group, called $su(2)$, we can choose for free 3 any independent traceless (trace equal to zero) matrices. The generators of any Lie group satisfy some algebraic and important relations. In the case of dealing with matrix or operator groups, the generators are matrices or operator themselves. These mathematical relations can be written in terms of (ordinary) algebraic commutators.

Feynman, Vernon, and Hellwarth developed geometric analogue of Rabi solutions, for two-level system, which provided additional physical insights. Such approach has been extended into $N$ level system. In this description the dynamics of  $N$ level system is visualised, without introducing new physics, as a vector in an $N^2-1$-dimensional vector space. This requires use of the $SU(N)$ symmetry to embed the atomic variables in the form of a vector with $N^2-1$ components. The exploitation of Lie algebra for study of dynamics has been extended into $N$-level system by Alhassid-Levine (AL)\cite{alhassid1977entropy,AlhassidLevine} and Hioe-Eberly(HE)\cite{Hioe1981,hioe1982nonlinear,hioe1983dynamic}. In Hioe-Eberly approach, for instance, the state of a three-level system is represented by pseudo-spin vector having 8 real components~(9 if we do not impose Normalization). This description provides an elegant geometrical framework.

The Quantum dynamics are calculated by the density matrix formalism \cite{fano1957description}. In this method  a statistical average over the assembly of the expectation values for the individual ions is included. This in turn yields information about observable macroscopic quantities. The formalism is advantageous in that it greatly simplifies the calculations since it allows one to handle a large number of dynamical variables in a systematic way. The physical significance of the individual density matrix elements depends on the representation with which the matrix is calculated. In a representation where $H_0$ is diagonal, the elements of the density matrix are defined as $\rho_{nm}=\langle\psi_n|\hat\rho|\psi_m\rangle$  where $\hat\rho$ is the density operator.  Recall that the dynamical evolution of an $N$-level atomic system is known to be studied in terms of density matrix $\hat\rho$ which obeys the Liouville equation \cite{levine2011quantum}
\begin{equation}
\begin{aligned}
i\hbar\frac{\partial}{\partial t}\hat\rho=&\left[\hat\rho,\hat H\right]
\label{Liouville}
\end{aligned}
\end{equation}
As by definition $\langle\hat O_{mn}\left(t\right)\rangle=Tr\left(\hat O_{mn} \hat\rho\left(t\right)\right)=\rho_{nm}\left(t\right)$ this implies that the dynamics can be equally investigated by studying $\hat O_{mn}$, where $\hat O_{mn}=|m\rangle\langle n|$. 

Although both paths, AL and HE, lead to same equation of motion for the expectation value of the generators the similarity is not straight forward because of the notation used in both formalism. While the AL method expressed closed Lie algebra as a commutation between the Hamiltonian and the generators, and in HE the closure is given by writing the Hamiltonian as a linear combination of the generators.  Moreover the equations of motion, for the expectation values of the generators, are given for row matrices in AL formalism where as HE provided the equation of motion for the columns. On top of this the background and motivation of both approaches also different. We here present the connection between these two derivation. The Alhassid-Levine formalism (AL) is developed in connection to surprisal analysis while the Hioe-Eberly (HE) aimed to extend the optical Bloch equations to three and more level systems. Optical Bloch equations of two level system are derived using both approaches as an example to complete the discussion. The use of expectation values, as opposed to operators, in deriving equation of motion has advantage in that it, i.e the average value, is a number that is measurable and easily solvable numerically and at times analytically. All this seem to make the two routes as if unconnected. This paper not only  aim to bring this two derivations together but also provide appropriate link which serves as a route from one formalism to another.

This work presents the connection between the two formalism, AL and HE.  To describe the equation of motion of the dynamics with the aid of Lie algebra firstly the atomic variables of a quantum  system are  embedded via $SU(N)$  Lie group yielding a new combination of group.  With this Lie group we construct a vector with $N^2-1$ components ($N$ components if normalization is not taken in to account), for $N$-level quantum system, to obtain the equation of motion describing the $SU(N)$ dynamics. Such approach has been shown to reveal hidden constants of motion \cite{hioe1982nonlinear,hioe1983dynamic}.  

The outline of the paper is as follows:  summary of the AL \& HE formalisms will be provided in Sec~(\ref{sec:ALHE}) before pointing out the link between them, AL and HE, in Sec~(\ref{sec:ReALHE}) following by derivation of the equation of motion in Sec~(\ref{sec:EQOM}) for two level system. Next in Sec~(\ref{sec:constmtn}) we provide discussion of constants of motion and super-evolution operator, before we provide concluding remarks in Sec~(\ref{sec:conclusion}).
\section{$SU(N)$ Dynamics}
\label{sec:ALHED}
In this section summary of derivations of both approach will be outlined. Both formalisms make use of the Liouville equation of motion to obtain the equation of motion for the expectation value of the generators. However AL use the commutation relation between the generators and the Hamiltonian while HE expanded the Hamiltonian as a linear combination of the generators. It is to be recalled here that every $N\times N$ density matrix is expressible in terms of a diagonal matrix and an element of of the Special unitary group $SU(N)$, generally speaking this provides $N^2$ real parameters if we do not impose normalization. 
\subsection{Summary of Alhassid-Levine (AL) and Hioe-Eberly (HE) formalisms}
\label{sec:ALHE}
For a $N$ level quantum system, where the Hilbert space is spanned by orthonormal states $|m\rangle,|k\rangle\ldots |n\rangle$. One readily sees that the $N^2$ operators $|i\rangle\langle j|, ~ i,j=1,2 \ldots N^2$ form a complete basis for the linear operators of the system. Using which any operator of the system can is expressible as a linear combination of these $N^2$ operators, with which we can form the desired Lie group, by embedding the atomic variables in them. To this end the Lie group used in both formalisms, i.e the generators $\hat G_\alpha$ (formed via linear combination of the $N^2$ operators), are traceless, orthogonal, Hermitian and obey the following properties \cite{Hioe1981,alhassid1977entropy}
\begin{subequations}
\begin{align}
Tr\left(\hat G_\alpha\hat G_\beta\right)=& 2\delta_{\alpha\beta}\label{TrGs}\\
\left[\hat G_{\alpha}, \hat G_{\beta}\right]=&2i f_{\alpha\beta\gamma} \hat G_\gamma\label{Gcommut}
\end{align}
\label{Gprop}
\end{subequations}
where $\delta_{\alpha\beta}$ is the usual Kronecker delta and $f_{\alpha\beta\gamma}$ is the structure constant which are antisymmetric in all indices. With which the density matrix as well as the Hamiltonian are written as a linear combination of the group generators \cite{Hioe1981,alhassid1977entropy}.
\begin{subequations}
\begin{align}
\hat\rho\left(t\right)=&\frac{\hat I}{N}+\frac{1}{2}\sum_{\alpha=1}^{N^2-1} \langle G_\alpha\rangle\left(t\right)\hat G_\alpha\label{expndRho}\\
\hat H\left(t\right)=&\frac{\hbar}{2}\left[\left(\sum_{\beta}^N\omega_\beta\right)\hat I+\sum_{\alpha=1}^{N^2-1} \Gamma_\alpha\left(t\right)\hat G_\alpha\right]\label{expndHaml}
\end{align}
\end{subequations}
where $\hbar\omega_\beta$ is energy of level $\beta$ and $\hat I$ is the identity operator. The coefficients $\langle G_\alpha\rangle\left(t\right)$ and $\Gamma_\alpha\left(t\right)$, expectation value of the generators and torque --respectively, are given by \cite{Hioe1981,alhassid1977entropy}
\begin{subequations}
\begin{align}
\langle G_\alpha\rangle\left(t\right)=&Tr\left(\hat\rho\left(t\right)\hat G_\alpha\right)\label{expectS}\\
\hbar\Gamma_\alpha\left(t\right)=&Tr\left(\hat H\left(t\right)\hat G_\alpha\right)\label{Torque}
\end{align}
\end{subequations}
Making use of the tracelessness of the generators  along with the  the expansion provided in Eq. \eqref{expndRho} it is to be noted that the trace of the density matrix  is $1$, $Tr(\rho)=1$
In their pioneering work on information theory Alhassid-Levine developed equation of motion for the expectation value of generators the summary of the derivation will be outlined. In the aforementioned scheme \cite{alhassid1977entropy} beginning from the expectation value of an operator, i.e $\langle G_{\alpha}\rangle\left(t\right)=Tr\left(\hat\rho\left(t\right)\hat G_\alpha\right)$, and taking derivative of both sides with respect to time, and making use of the Liouville equation of motion for density matrix one arrives at
\begin{equation}
\begin{aligned}
\frac{d}{dt}\langle G_{\alpha}\rangle=& Tr\Big(\frac{1}{i\hbar}[\hat H,\hat\rho]\hat G_{\alpha}\Big)
\label{vecG4AL0}
\end{aligned}
\end{equation}
Using cyclic property of trace and noting that the Hamiltonian is closed under Lie algebra with the generators
\begin{equation}
\begin{aligned}
\left[\hat H,\hat G_\alpha\right]=i \hbar\sum_\beta \hat G_\beta g_{\beta\alpha}
\end{aligned}
\label{HcomGAL}
\end{equation} 
one readily obtains the  equation of motion for the expectation value of generators,  given to be 
\begin{equation}
\begin{aligned}
\frac{d}{dt}\langle G_{\alpha}\rangle=&-\sum_{\beta}\langle G_{\beta}\rangle g_{\beta\alpha}
\label{vecG4AL}
\end{aligned}
\end{equation}
where $\alpha,~\beta=1,2,\dots,N^2-1$, $\langle G_{\beta}\rangle$ is row, and the $g_{\beta\alpha}$ are (possibly complex) numerical coefficients. 

On the other hand, in HE formalism  the Liouville equation of motion for the density matrix is multiplied, on both sides by generators and then trace, $Tr$, is taken to yield \cite{Hioe1981} :
\begin{equation}
\begin{aligned}
i\hbar\frac{d}{dt}\langle G_{\alpha}\rangle=& Tr\Big([\hat H,\hat\rho]\hat G_{\alpha}\Big)\\
=& Tr\Big(\frac{\hbar}{2}\Gamma_{\beta}\left(t\right)[\hat G_{\alpha},\hat G_{\beta}]\hat\rho\Big)
\label{vecG4HE0}
\end{aligned}
\end{equation}
Where the cyclic property of trace and expansion of Hamiltonian as given in Eq.\eqref{expndHaml} has been made use of. Note also that the Hamiltonian commutes with Identity. Next, with the aid of the commutation relation Eq.\eqref{Gcommut} and \eqref{expectS}, i.e the expectation value of the generators, one readily arrives at the equation of motion for the expectation values of the generators 
\begin{equation}
\begin{aligned}
\frac{d}{dt}\langle G_{\alpha}\rangle=&\Gamma_{\beta}\left(t\right)f_{\alpha\beta\gamma}\langle G_{\gamma}\rangle 
\label{vecG4HE}
\end{aligned}
\end{equation}
Therefore as can be seen from,  Eq.\eqref{vecG4AL} and Eq.\eqref{vecG4HE}, of AL and HE respectively, at glance seems unrelated as the notation used are slightly different. It is the intention of the next section to establish the link between these two routes. 
\subsection{Connection between Alhassid-Levine (AL) and Hioe-Eberly (HE) formalisms}                                                                             
\label{sec:ReALHE}
In the AL formalism, Eq. \eqref{vecG4AL} is equation of motion for the generator where $\langle G_{\beta}\rangle$ is row. So to make comparison with HE formalism we need to write them as equation of motions for column $\langle G_{\beta}\rangle^T$, in doing so we should note that due to antisymmetric nature of the numerical coefficients the transpose of the $g$ matrix takes the form $g_{\alpha\beta}=-g_{\beta\alpha}$
\begin{equation}
\begin{aligned}
\frac{d}{dt}\langle G_{\alpha}\rangle=&g_{\alpha\beta} \langle G_{\beta}\rangle 
\label{vecG4ALc}
\end{aligned}
\end{equation}
where indices represent sum over. The comparison of Eq. \eqref{vecG4ALc} with that of HE, Eq. \eqref{vecG4HE}, yields that
\begin{equation}
\begin{aligned}
g_{\beta\alpha}=&\Gamma_{\gamma}\left(t\right) f_{\gamma\alpha\beta}
\label{AL2HE}
\end{aligned}
\end{equation}
Thus the matrix $g$ of AL is the product between the torque vector and antisymmetric structure constant $f_{\alpha\beta\gamma}$ of HE.
\section{Example: Equation of motions for two level system }                                                                             
\label{sec:EQOM}
We now turn to the physical application of the $SU(2)$  groups and $su(2)$ algebras. To see example of the approach we here take a simple  and familiar two level system and describe the dynamics using $SU(2)$ group. The two level system we take is very well studied. We consider a two level system  where  a laser pulse perturbs the systems. For the purpose at hand all we need to know is its Hamiltonian to obtain the equation of motion for the expectation value of the generators. The Hamiltonian we consider is under Rotating Wave Approximation (RWA) and in the interaction picture is known to be \cite{Bergmann2001,shore2008}.
\begin{equation}
\hat H \left(t\right)=\frac{\hbar}{2}\begin{pmatrix}
0 & \Omega  \left(t\right)\\
\Omega \left(t\right) & 2\Delta \\
 \end{pmatrix} 
\end{equation}
where $\frac{\hbar\Omega \left(t\right)}{2}$ is  half-Rabi frequency of the pulse applied and $\hbar\Delta$, i.e detuning, is the difference between the laser frequency and Bohr frequency.  

 For a two-level quantum system, where the Hilbert space is spanned by two orthonormal states $|0\rangle$ and $|1\rangle$. One readily sees that the four operators $|i\rangle\langle j|, ~ i,j=0,1$ form a complete basis for the linear operators of the system in the sense that any linear operator of the system can be written as a linear superposition (with complex coefficients which can be time dependent) of these four operators. It therefore is possible to chose the Pauli matrices, along with the identity operator $\hat I=|0\rangle\langle 0|+|1\rangle\langle 1|$, as our generators, and they are given by \cite{MichaelChuang2010, Eberly1975}
\begin{equation}
\begin{aligned}
 \hat G_1=\begin{pmatrix}
0 & 1\\
1 & 0 \end{pmatrix},       & & \hat G_2=\begin{pmatrix}
0 & -i\\
i & 0 \end{pmatrix} ,      & & \hat G_3=\begin{pmatrix}
1 & 0\\
0 & -1 
 \end{pmatrix}  
 \end{aligned}
 \label{gene}
\end{equation}
The structure constant $f_{\alpha\beta\gamma}$ in this case is the Levi-Civita $\epsilon_{\alpha\beta\gamma}$.  Moreover the non-vanishing components of the fully antisymmetric tensor $\epsilon$ of the constants of $SU(2)$ group given by 
\begin{equation}
\begin{aligned}
\epsilon_{123}=&\epsilon_{231}=\epsilon_{312}=1\\
\epsilon_{132}=&\epsilon_{213}=\epsilon_{321}=-1
\end{aligned}
\label{structSU2}
\end{equation}
To explore the dynamics of the two level using the both scheme, we here express our Hamiltonian, to within an addition of multiple of an identity matrix which commutes with the Hamiltonian, in terms of the generators of the $SU (2)$ group Eq.\eqref{gene} as \cite{Cahn1984} 
\begin{equation}
\begin{aligned}
\hat H\left(t\right)=\frac{\hbar}{2}\Big[\Omega\left(t \right)\hat G_1-\Delta\hat G_3\Big]+const
\end{aligned}
\label{ham2}
\end{equation} 
where $const=\frac{\hbar}{2}\Big(\sum_{\beta=1}^2\omega_\beta\Big)\hat I$, $\hbar\omega_\beta$ is energy of level $\beta$. Note here that, with the exception of a term proportional to $\hat I$, which is constant and thus commutes with any linear operator of the system, the Hamiltonian is a linear combination with real coefficients of the three $su(2)$ generators. This means that the dynamical algebra of the two-level atomic system is $su(2)$. 

Next the equation of motion of a coherence vector whose elements  are the expectation value of the generators as given by Eq.\eqref{expectS} formed to be  $\vec G=\left(\langle G_1\rangle\left(t\right),\langle G_2\rangle\left(t\right),\langle G_3\rangle\left(t\right)\right)^T$ is obtained.

In the AL formalism one is able to find the matrix elements $g_{\beta\alpha}$ using the commutation relation in Eq.\eqref{HcomGAL} and Eq.\eqref{ham2} along with the commutation relation between the generators. For instance we note that the commutation between the Hamiltonian and the generator $\hat G_2$ is 
\begin{equation}
\begin{aligned}
\left[\hat H\left(t\right), \hat G_{2}\right]=& i \hbar \hat G_\beta g_{\beta 2}\\
i \hbar \left( \Omega\left(t\right) \hat G_{3} + \Delta\hat G_{1}\right)=& i \hbar \hat G_\beta g_{\beta 2}\\
\end{aligned}
\label{HcomG2}
\end{equation}
where the indices means sum over. From the last line of Eq.\eqref{HcomG2} it follows that 
\begin{equation}
\begin{aligned}
g_{32}=&\Omega\left(t\right),  && g_{12}=\Delta
\end{aligned}
\label{gx2}
\end{equation}
In like manner one is able to find all the $g_{\beta\alpha}$. Using this knowledge and Eq.\eqref{vecG4AL} the equation of motion, say for $\langle \dot G_2\rangle$, is obtained as
\begin{equation}
\begin{aligned}
\frac{d}{dt}\langle G_2\rangle\left(t\right)=&-\langle G_\beta \rangle g_{\beta2}\\
=&-\Delta\langle G_1 \rangle -\Omega\left(t\right)\langle G_3 \rangle
\end{aligned}
\label{dG2dt}
\end{equation}
To find the equation of motion using HE formalism, we note the torque vector to be $\vec\Gamma\left(t\right)=(\Omega\left(t\right), 0,  -\Delta)^T$, equipped with this and Eq.\eqref{vecG4HE} it is possible to reproduce the previous result as follows
\begin{equation}
\begin{aligned}
\frac{d}{dt}\langle G_2\rangle\left(t\right)=&\Gamma_\beta\left(t\right)f_{2\beta\gamma}\langle G_\gamma \rangle\\
=&-\Delta\langle G_1 \rangle -\Omega\left(t\right)\langle G_3 \rangle
\end{aligned}
\label{dG2dtHE}
\end{equation}

Hence using either of the approach with as given in Eq.\eqref{vecG4AL} and Eq.\eqref{vecG4HE}  the  equation of motion for the coherence vector can readily obtained. 
\begin{equation}
\begin{aligned}
\frac{d}{dt}\vec G=&g\vec G
\label{sdot}
\end{aligned}
\end{equation}
 where $g$ is a $3\times 3$ antisymmetric matrix given by 
\begin{equation}
\begin{aligned}
g=&
 \begin{pmatrix}
  0 & \Delta  & 0 \\
  -\Delta & 0 & -\Omega\left(t\right) \\
  0 & \Omega\left(t\right) & 0
 \end{pmatrix}
\end{aligned}
\label{matrixg}
\end{equation} 

\section{Constants of Motion}
\label{sec:constmtn}
States are labeled by specific values of their properties, which do not change with time - these properties are called constants of motion. A constant of motion in quantum mechanics is defined as an observable-operator that commutes with the Hamiltonian. In two-level atoms (and also in N-level atoms), constants of motion are conveniently discussed in terms of the Bloch sphere and restrictions for the Bloch vector moving on the Bloch sphere. In this case, the underlying symmetry group that can explain the constants of motion is $SU(N)$. The Hamiltonian of the N-level atom can be expressed in terms of the group generators of $SU(N)$, and the dynamical variables of the system can be associated with these generators. Hioe and Eberly have shown how to exploit the properties of the $su(N)$ algebra in the formulation of the equations of motion of the dynamical variables of an N-level system, and how one can obtain conservation laws (constants of motion) corresponding to $SU(N)$. 

In the two-level system the generator $\hat I$ commutes with the rest of the generators, i.e $[\hat G_\alpha,\hat I]=0$ -- where $\alpha=1,2,3$. This in turn implies that the generator $\hat I$  does not contribute to the dynamical equations of motion, as it commutes with any linear operator of the system, this means that the generator $\hat I$  is a constant of motion (i.e. $\frac{d\hat I}{dt}=0$). Moreover in search of constants of motion beyond the identity operator let us now assume that the detuning to be time dependent. Furthermore consider the case where both the detuning and the laser pulse to have same time dependence but  different amplitude.
\begin{equation}
\begin{aligned}
\Omega\left(t\right)=&\Omega_0 q\left(t\right)\\
\Delta\left(t\right)=&\Delta_0 q\left(t\right)
\end{aligned}
\label{DeltaOmega}
\end{equation}
with $q\left(t\right), \Omega_0, \Delta_0$ the time dependence function,  peak values of the laser pulse, and peak value of the Detuning respectively. Introducing $\epsilon\left(t \right)=\sqrt{\Omega^2+\Delta^2}$ and thus following footsteps of \cite{PhysRevA.11.1641} this would equip us with dimensionless parameters $\frac{\Omega}{\epsilon},\frac{\Delta}{\epsilon}$. Armed with this new information let us now make transformation of the generators $\hat G_\alpha$, of Eq.\eqref{gene}, into new generators labelled as $\hat F_\alpha$
\begin{equation}
\begin{aligned}
\hat F_1=&\frac{\Omega}{\epsilon} \hat G_1-\frac{\Delta}{\epsilon} \hat G_3\\
\hat F_2=&\hat G_2\\
\hat F_3=&\frac{\Delta}{\epsilon} \hat G_1-\frac{\Omega}{\epsilon} \hat G_3
\end{aligned}
\label{Fgene}
\end{equation}
Note here that this new generators have same commutation relation as the generators $\hat G_\alpha$. Also notably, using Eqs. \eqref{expndRho} and \eqref{expndHaml}, we can express the density matrix and the Hamiltonian in terms of the new generators, and consequently the Hamiltonian takes the form
\begin{equation}
\begin{aligned}
\hat H\left(t\right)=\hbar\epsilon\left(t \right)\hat F_1+const
\end{aligned}
\label{hamF2}
\end{equation} 
Equation \eqref{hamF2} reveals two interesting and points, first notice the Hamiltonian is no more dependent on the generators $\hat F_3$, compare with Eq.\eqref{ham2}, because the Hamiltonian now commutes with the generator $\hat F_1$, i.e. $[\hat H\left(t\right),\hat F_1]=0$, we conclude $\hat F_1$ is a constant of motion and it thus follows that
 \begin{equation}
\begin{aligned}
\frac{d}{dt}\hat F_1=0
\end{aligned}
\label{dF1dt}
\end{equation} 
Following the steps outline in section \ref{sec:EQOM}, and using the notation $\vec F=(F_1,F_2,F_3)^T$, where $F_\alpha$ is the expectation value of the generators, one readily obtains the dynamical equations of motion for the new generators to be
\begin{equation}
\begin{aligned}
\frac{d}{dt}\vec F=&f\vec F
\end{aligned}
\label{dFdt1}
\end{equation}  
where 
\begin{equation}
\begin{aligned}
f=&\begin{pmatrix}
f_{1\times 1} & 0\\
0  & f_{2\times2}\end{pmatrix}, \\
f_{1\times 1}=&\begin{pmatrix} 0 \end{pmatrix}, &
f_{2\times2}=&\begin{pmatrix}
0 & -\epsilon\\
\epsilon  & 0\end{pmatrix}
\end{aligned}
\label{dFdt2}
\end{equation} 
This informs us that the newly transformed vector $\vec F$ is decomposed into two vectors that live in separate subspaces  and consequentially yields two more constants of motion
\begin{equation}
\begin{aligned}
F_1\left(t\right)^2=&const,\\
F_2\left(t\right)^2+F_3\left(t\right)^2=& const
\end{aligned}
\label{dFdtconst}
\end{equation} 
Physically the last equation means that the length of $\vec F_1$ and $\vec F_{23}=(F_2,  F_3)^T$ is separately conserved, in addition to this, as usual, the total length of the vector $\vec F$ is also conserved~\cite{Eberly1975,hioe1982nonlinear,hioe1983dynamic},
\begin{equation}
\begin{aligned}
F_1\left(t\right)^2+F_2\left(t\right)^2+F_3\left(t\right)^2=& const
\end{aligned}
\label{Fdtconst}
\end{equation} 
It is worth pointing out that for exact resonance, i.e. $\Delta=0$, one readily notes that $\hat F_\alpha=\hat G_\alpha$ and because now $[H, G_1]=0$ at exact resonance $G_1$ is new constant of motion. Setting $\Delta=0$ shows that now the dynamical equation of motion can be decomposed into  two subspaces each of which yield constants of motion given to be
\begin{equation}
\begin{aligned}
G_1\left(t\right)^2=&const,\\
G_2\left(t\right)^2+G_3\left(t\right)^2=& const
\end{aligned}
\label{dgdtconst}
\end{equation} 
From which follows that the length of $\vec G_1$ and $\vec G_{23}=( G_2, G_3)^T$ is separately conserved, in addition to this, as usual, the total length of the vector $\vec G$ is also conserved~\cite{Eberly1975}. 

Another important feature to notice from Eq.\eqref{dFdt1} is that, not only the block diagonal matrix $f\left(t\right)$ commutes with itself at different times, i.e $[f\left(t_1\right),f\left(t_2\right)]=0$, but also the disjoint matrices $f_k\left(t\right), k=1,2$ commutes with thierself at different times, i.e $[f_k\left(t_1\right),f_k\left(t_2\right)]=0$. This means that we can employe the Magnus solution, outlined in section~(\ref{sec:Magnus}), and  obtain the solution for the new vector to be $\vec{F}\left(t\right)=M\left(t,0\right)\vec{F}\left(0\right)$ where $M\left(t,0\right)$ is given to be
\begin{equation}
\begin{aligned}
M\left(t,0\right)=&\begin{pmatrix}
1 & 0 &0\\
0 & \cos\epsilon' &-\sin\epsilon'\\
0 & \sin\epsilon' & \cos\epsilon'
\end{pmatrix}
\end{aligned}
\label{MstF1}
\end{equation}
 where $\epsilon'=\int_{0}^t\epsilon\left(t'\right)dt'$. To this end iff we again assume the system is initially prepared to be on the ground state, this entails to the initial vector being $\vec F\left(0\right)=(-\frac{\Delta_0}{\epsilon_0},0,\frac{\Omega_0}{\epsilon_0})^T$ with $\epsilon_0=\sqrt{\Omega_0^2+\Delta_0^2}$ hence the solution takes the form
\begin{equation}
\begin{aligned}
\vec{F}\left(t\right)=&
 \left(
 -\frac{\Delta_0}{\epsilon_0},
  -\frac{\Omega_0\sin{\epsilon'}}{\epsilon_0},
 \frac{\Omega_0\cos{\epsilon'}}{\epsilon_0}
 \right)^T
\end{aligned}
\label{stF1}
\end{equation}
We can see from the solution Eq.\eqref{stF1} that $F_1\left(t\right)=F_1\left(0\right)$ does not change with time and 
\begin{equation}
\begin{aligned}
|F_1\left(t\right)|^2=&|F_1\left(0\right)|^2\\
|F_2\left(t\right)|^2+|F_3\left(t\right)|^2=&|F_2\left(0\right)|^2+|F_3\left(0\right)|^2
\end{aligned}
\label{FkConserve}
\end{equation} 
To conclude we showed that the time evolution of the pseudospin vector $\vec G$ can be studied in terms of the time evolution of two independent vectors of dimensions one and two, evolving in two separate subspaces of their respective dimensions provided the conditions of Eq.\eqref{DeltaOmega} are met; and/or if system under consideration is on exact resonance (i.e. $\Delta=0$). Following this fact we then provided two constants of motion representing the square of the length of each vector. Moreover the sum of these two vectors is the square of the length of the vector which is commonly known as the constant of motion $\vec G$.
\subsection{Super-Evolution Matrix }
\label{sec:supeEvolWN}
In this section we will explore the super-evolution, where super stands for an operator acting on another operator, matrix that will be obtained via Wei-Norman approach, WN \cite{wei1963lie,wilcox1967exponential}. WN technique tells us that the evolution operator, of a Hamiltonian that is a linear combination of Lie group generators, can be expressed as an ordered product of exponentials, whose arguments are the product of a time-dependent
function and generator of the group. 
Solution of Eq.\eqref{dFdt1} can be formally written as 
\begin{equation}
\begin{aligned}
\vec F\left(t\right)=&M\big(t,0\big)\vec F\left(0\right)
\end{aligned}
\label{transform}
\end{equation}
where $\vec F$ being three dimensional column vector  whose components are the average values of the observables, and the super-evolution matrix $M$ an $3\times 3$ matrix. 
At this point it is helpful relating  the super-evolution matrix $M$ to the evolution operator $\hat U$. To this aim, let us notice that, according to the $3\times 3$ representation of the generators $\hat F_{\alpha}$,  \cite{dattoli1991matrix}
\begin{equation}
\begin{aligned}
\big(\digamma_{\alpha}\big)_{\beta\gamma}=&-\imath f_{\alpha,\beta,\gamma}
\end{aligned}
\label{adjF}
\end{equation}
This representation is called adjoint representation. The explicit matrix elements of the adjoint representation are given by the structure constants of the algebra. One can rewrite the Liouville equation in the following form (for $\hbar=1$)
\begin{equation}
\begin{aligned}
\imath\frac{d}{dt}\rho\left (t\right)=&[H,\rho]\\
\imath\frac{d}{dt}\rho\left (t\right)=&ad_H\left(\rho\right)
\end{aligned}
\label{adjL}
\end{equation}
where $ad_H$ is so-called commutator superoperator, because it is an operator acting on another operator, and is defined to be
\begin{equation}
\begin{aligned}
ad_H=\frac{\hbar}{2}\sum_{\alpha}\Gamma_{\alpha}\left(t\right)ad_{F_{\alpha}}
\end{aligned}
\label{adjH}
\end{equation}
where $ad_{F_{\alpha}}$ is defined, following Eq.\eqref{adjF} to be
\begin{equation}
\begin{aligned}
\big(ad_{F_{\alpha}}\big)_{\beta\gamma}=&-\imath f_{\alpha,\beta,\gamma}
\end{aligned}
\label{adjFalpha}
\end{equation}
On making use of the adjoint representation and the fact that $\langle \mathcal F\left(t\right)\rangle =Tr\big(\hat\rho\left(t\right)\hat  {\mathcal F}\big)$ we now can rewrite the equation of motion for the coherence vector in the following form 
\begin{equation}
\begin{aligned}
\imath\frac{d}{dt}\vec {\mathcal F}=&ad_H\vec  {\mathcal F}
\end{aligned}
\label{adjFch}
\end{equation}
Notice that the passing from $\rho$ in Eq.\eqref{adjL} to $\mathcal F$ in Eq.\eqref{adjFch} is equivalent to passing to the adjoint representation of the Lie algebra $su(N)$. If we now use the notation $ad_H=\mathcal H$ and $ad_{F_{\alpha}}=\mathcal F$ we can rewrite Eq.\eqref{adjFch}  in the matrix form as
\begin{equation}
\begin{aligned}
\imath \frac{d}{dt} \mathcal F=&\mathcal H \mathcal F
\label{adjdiF}
\end{aligned}
\end{equation}
Using Eq.\eqref{transform} it turns into the Schr\"{o}dinger equation for the matrix $M\left(t,0\right)$
\begin{equation}
\begin{aligned}
i\frac {d }{dt}M\left(t,0\right)=&\mathcal H M\left(t,0\right), &&M\left(t,0\right)=I\\
\end{aligned}
\end{equation}
Dropping the time argument $M\left(t,0\right)=M$ for simplicity. 
\begin{equation}
\begin{aligned}
i\frac {d M}{dt}M^{-1}=&\mathcal H ,
\end{aligned}
\label{eqMtnM}
\end{equation}
The interpretation of $M$ as the geometrical counterpart of $\hat U$ immediately follows. In the WN approach the evolution matrix is expressed in terms of closed lie algebra; for generators $\hat  {\mathcal F}$ with corresponding time dependent coefficients $\Upsilon\left(t\right)$, the unitary super-evolution is expressed as 
\begin{equation}
\begin{aligned}
\hat M \left(t\right)=&\prod_{\alpha}e^{\Upsilon_{\alpha}\left(t\right)\hat  {\mathcal F}_{\alpha}}\\
\end{aligned}
\label{weiNrmnM}
\end{equation}
where the adjoint representation generators $ {\mathcal F_\alpha}$
\begin{equation}
\begin{aligned}
\mathcal F_1=&\begin{pmatrix}
0 & 0 & 0\\
0 & 0 & -1\\
 0 & 1 & 0 \end{pmatrix}, &&
 \mathcal F_2=&\begin{pmatrix}
0 & 0 & 1\\
0 & 0 & 0\\
 -1 & 0 & 0 \end{pmatrix}, &&
 \mathcal F_3=&\begin{pmatrix}
0 & -1 & 0\\
1 & 0 & 0\\
 0 & 0 & 0 \end{pmatrix} 
\end{aligned}
\label{mathCalF}
\end{equation}
From Eqs.\eqref{eqMtnM}, \eqref{weiNrmnM} and the adjoint generators \eqref{mathCalF} we 
\begin{equation}
\begin{aligned}
\frac {d M}{dt}M^{-1}=&\dot\Upsilon_1\mathcal F_1+\dot\Upsilon_2~e^{\Upsilon_1\mathcal F_1}\mathcal F_2~e^{-\Upsilon_1\mathcal F_1}+\dot\Upsilon_3~e^{\Upsilon_1\mathcal F_1}~e^{\Upsilon_2\mathcal F_2}\mathcal F_3~e^{-\Upsilon_2\mathcal F_2}~e^{-\Upsilon_1\mathcal F_1}
\end{aligned}
\label{eqMtnMc}
\end{equation}
Using the Baker-Campell-Hausrdoff relation 
\begin{equation}
\begin{aligned}
e^A~B~e^{-A}=& B + [A,B]+\frac{1}{2!}[A,[A,B]]+\frac{1}{3!}[A,[A,[A,B]]]+\ldots
\end{aligned}
\label{BCH}
\end{equation}
we readily see that
\begin{equation}
\begin{aligned}
e^{\Upsilon_1\mathcal F_1}\mathcal F_2~e^{-\Upsilon_1\mathcal F_1}=& \mathcal F_2\cos\Upsilon_1+\mathcal F_3\sin\Upsilon_1\\
e^{\Upsilon_1\mathcal F_1}e^{\Upsilon_2\mathcal F_2}\mathcal F_3~e^{-\Upsilon_2\mathcal F_2}~e^{-\Upsilon_1\mathcal F_1}=&\mathcal F_1\sin\Upsilon_2-\mathcal F_2\cos\Upsilon_2\sin\Upsilon_1+\mathcal F_3\cos\Upsilon_2\cos\Upsilon_1
\end{aligned}
\label{BCHre}
\end{equation}
Equating with the Hamiltonian, for the Hamiltonian given in Eq.\eqref{matrixg}, and putting it in matrix form we get (note the Hamiltonian becomes $\mathcal H=\epsilon\mathcal F$ and the RHS will be $(\epsilon,0,0)^T$ for the case given by Eq.\eqref{dFdt1})
\begin{equation}
\begin{aligned}
\begin{pmatrix}
1 & 0 & \sin\Upsilon_2\\
0 & \cos\Upsilon_1 & -\cos\Upsilon_2\sin\Upsilon_1\\
 0 & \sin\Upsilon_1 & \cos\Upsilon_1\cos\Upsilon_2 \end{pmatrix} \begin{pmatrix}
\dot\Upsilon_1\\
\dot\Upsilon_2\\
 \dot\Upsilon_3 \end{pmatrix}=&\begin{pmatrix}
\Omega\\
 0\\
 -\Delta \end{pmatrix} 
\end{aligned}
\label{wenMat}
\end{equation}
In short hand notation we cab write Eq.\eqref{wenMat} as
\begin{equation}
\begin{aligned}
W\dot\Upsilon=&I
\end{aligned}
\label{wenMatshrt}
\end{equation}
So as to obtain solution we demand that $Det |W|=\cos\Upsilon_2\neq0$, and the inverse of $W$ takes the form
\begin{equation}
\begin{aligned}
W^{-1}=&
\begin{pmatrix}
1 & \sin\Upsilon_1\tan\Upsilon_2 & -\cos\Upsilon_1\tan\Upsilon_2\\
0 & \cos\Upsilon_1 & \sin\Upsilon_1\\
 0 & \sec\Upsilon_1\sin\Upsilon_2 & \cos\Upsilon_1\sec\Upsilon_2 \end{pmatrix} 
 \end{aligned}
\label{inverseW}
\end{equation}
From Eq.\eqref{wenMat} we get that
\begin{equation}
\begin{aligned}
\dot\Upsilon_1+\dot\Upsilon_3\sin\Upsilon_2=\Omega\\
\dot\Upsilon_2\cos\Upsilon_1-\dot\Upsilon_3\cos\Upsilon_2\sin\Upsilon_1=0\\
\dot\Upsilon_2\sin\Upsilon_1+\dot\Upsilon_3\cos\Upsilon_2\cos\Upsilon_1=-\Delta\\
\end{aligned}
\label{param}
\end{equation}
which yields
\begin{equation}
\begin{aligned}
\Upsilon_1=&\cot^{-1}\left(\cos\Upsilon_2\frac{\dot\Upsilon_3}{\dot\Upsilon_2}\right)\\
\Delta=&\sqrt{\dot\Upsilon_2^2+\dot\Upsilon_3^2\cos^2\Upsilon_2}\\
\Omega=&\dot\Upsilon_1+\dot\Upsilon_3\sin\Upsilon_2\\
\end{aligned}
\label{paramsol}
\end{equation}
For the case with the block diagonal Hamiltonian $f$ with $\mathcal H=\epsilon\mathcal F$ where we have $(\epsilon,0,0)^T$ in the RHS of Eq.\eqref{wenMat}, the solution yields
\begin{equation}
\begin{aligned}
\dot\Upsilon_1=&\epsilon,  & & \dot\Upsilon_2=\dot\Upsilon_3=0
\end{aligned}
\label{paramfsol}
\end{equation}
The case $\Upsilon_2=0$ reproduces the evolution matrix Eq.\eqref{MstF1}, i.e $W=M$,  with $\Upsilon_1=\epsilon'=\int_{0}^t\epsilon\left(t'\right)dt'$
\begin{equation}
\begin{aligned}
W=&
\begin{pmatrix}
1 & 0 & 0\\
0 & \cos\Upsilon_1 & -\sin\Upsilon_1\\
 0 & \sin\Upsilon_1 & \cos\Upsilon_1 \end{pmatrix}
 \end{aligned}
\label{WeqlsM}
\end{equation}
\section{Conclusion} 
\label{sec:conclusion}
The Alhassid-Levine formalism (AL) is developed in connection to surprisal analysis while the Hioe-Eberly (HE) aimed to extend the optical Bloch equations to three and more level systems. Connection is made between these two formalisms that made use of Lie algebra to describe equation of motion for quantum system. In both approaches the Hamiltonian and the density matrix are expressed as a linear combination of the Lie group. While expansion of density matrix in terms of $SU(N)$ is true for all density matrices, the expansion of Hamiltonian in terms of the generators works only for special case of Hamiltonian. We provided a link that bridges the gap between these two formalism. With appropriate temporal profile for the laser pulse and detuning we showed how a new generator can be obtained which provides two constants of motion evolving independently on their own subspaces. Exact analytical solution (using Wei-Norman formalism)  of the optical Bloch equation is  also provided. 

\bibliography{mybibfile}

\end{document}